\documentstyle[12pt]{article}
 \begin{document} 


\newskip\humongous \humongous=0pt plus 1000pt minus 1000pt
\def\caja{\mathsurround=0pt}
\def\eqalign#1{\,\vcenter{\openup1\jot \caja
 \ialign{\strut \hfil$\displaystyle{##}$&$
 \displaystyle{{}##}$\hfil\crcr#1\crcr}}\,}
\newif\ifdtup
\def\panorama{\global\dtuptrue \openup1\jot \caja
 \everycr{\noalign{\ifdtup \global\dtupfalse
 \vskip-\lineskiplimit \vskip\normallineskiplimit
 \else \penalty\interdisplaylinepenalty \fi}}}
\def\eqalignno#1{\panorama \tabskip=\humongous
 \halign to\displaywidth{\hfil$\displaystyle{##}$
 \tabskip=0pt&$\displaystyle{{}##}$\hfil
 \tabskip=\humongous&\llap{$##$}\tabskip=0pt
 \crcr#1\crcr}}
\jot = 1.5ex
\def\baselinestretch{1.2}
\parskip 5pt plus 1pt
\catcode`\@=11
\@addtoreset{equation}{section}
\def\theequation{\arabic{section}.\arabic{equation}}
\def\@normalsize{\@setsize\normalsize{15pt}\xiipt\@xiipt
\abovedisplayskip 14pt plus3pt minus3pt%
\belowdisplayskip \abovedisplayskip
\abovedisplayshortskip \z@ plus3pt%
\belowdisplayshortskip 7pt plus3.5pt minus0pt}
\def\small{\@setsize\small{13.6pt}\xipt\@xipt
\abovedisplayskip 13pt plus3pt minus3pt%
\belowdisplayskip \abovedisplayskip
\abovedisplayshortskip \z@ plus3pt%
\belowdisplayshortskip 7pt plus3.5pt minus0pt
\def\@listi{\parsep 4.5pt plus 2pt minus 1pt
     \itemsep \parsep
     \topsep 9pt plus 3pt minus 3pt}}
\relax
\catcode`@=12
\evensidemargin 0.0in
\oddsidemargin 0.0in
\textwidth 6.0in
\textheight 8.5in
\hoffset .7 cm
\voffset -1 cm
\headsep .25in
\catcode`\@=11
\def\section{\@startsection{section}{1}{\z@}{3.5ex plus 1ex minus
   .2ex}{2.3ex plus .2ex}{\large\bf}}

\def\thesection{\arabic{section}}
\def\thesubsection{\arabic{section}.\arabic{subsection}}
\def\thesubsubsection{\arabic{subsubsection}.}
\def\appendix{\setcounter{section}{0}
 \def\thesection{Appendix \Alph{section}}
 \def\theequation{\Alph{section}.\arabic{equation}}}
\newcommand{\beq}{\begin{equation}}
\newcommand{\eeq}{\end{equation}}
\newcommand{\bea}{\begin{eqnarray}}
\newcommand{\eea}{\end{eqnarray}}
\newcommand{\beas}{\begin{eqnarray*}}
\newcommand{\eeas}{\end{eqnarray*}}
\newcommand{\defi}{\stackrel{\rm def}{=}}
\newcommand{\non}{\nonumber}
\def\de{\partial}
\def\si{\sigma}
\def\dim{\hbox{\rm dim}}
\def\sup{\hbox{\rm sup}}
\def\inf{\hbox{\rm inf}}
\def\Arg{\hbox{\rm Arg}}
\def\Im{\hbox{\rm Im}}
\def\Re{\hbox{\rm Re}}
\def\Res{\hbox{\rm Res}}
\def\Max{\hbox{\rm Max}}
\def\Abs{\hbox{\rm Abs}}
\def\infi{\infty}
\def\nrm{\parallel}

\def\e{\varepsilon}
\def\s{\sigma}
\def\cf{{\cal C}_F}
\def\tcf{{\left({\tau\over \cf}\right)}}

\begin{titlepage}
\begin{center}
{\Large
Tricritical Ising Model near criticality}
\end{center}
\vspace{1ex}

\begin{center}
{\large
Riccardo Guida$^{1}$ and Nicodemo Magnoli$^{2,3}$}
\end{center}
\vspace{1ex}
\begin{center}
{\it $^{1}$ CEA-Saclay, Service de Physique Th\'eorique\\
     F-91191 Gif-sur-Yvette Cedex, France}\\
{\it $^{2}$ Dipartimento di Fisica -- Universit\`a di Genova\\
     Via Dodecaneso, 33 -- 16146 Genova, Italy}\\ 
{\it $^{3}$ Istituto Nazionale di Fisica Nucleare- Sez. Genova\\
     Via Dodecaneso, 33 -- 16146 Genova, Italy}\\
\end{center}

\begin{center}
e-mail: guida@spht.saclay.cea.fr, magnoli@ge.infn.it
\end{center}
\medskip
{\bf ABSTRACT:}
The most relevant thermal perturbation of the
continuous $d=2$ minimal conformal theory with $c=7/10$
(Tricritical Ising Model) is treated here.
This model describes the scaling region of the 
$\phi^6$ universality class near the tricritical point.
The problematic IR divergences of the
naive perturbative expansion
around conformal theories are dealt within the
OPE approach developed at all orders by the authors.
The main result 
is a description of the short distance
behaviour of correlators that is compared with 
existing  
long distance expansion (form factors approach) related 
to the  integrability of the model.  

\bigskip
\begin{flushleft}
SPhT-t$96/142$ 

GEF-Th-$15$ 

Revised Version 4-1997

\newpage

{\bf Corresponding Author:} Riccardo Guida;\par
address: CEA-Saclay, Service de Physique Th\'eorique \par
         F-91191 Gif-sur-Yvette Cedex, France\par
email: guida@spht.saclay.cea.fr;\par
tel: 00 33 1 69088116\par
fax: 00 33 1 69088120.

\end{flushleft}
\end{titlepage}

\section{Introduction}\label{introduction}
The  $d=2$ minimal conformal quantum field theory
whose central charge is equal to $c=7/10$, 
known as Tricritical Ising Model (TIM),
describes 
the scaling region near the
tricritical point (i.e. the end point of a line where three phases
coexist simultaneously)
of the universality class 
corresponding to the  
Landau-Ginzburg theory with an interaction $\phi ^6$ \cite{zamotim}.
In particular it can be thought of as a continuum 
realization of the Ising model
with annealed vacancies, see  
\cite{blume}, and 
it describes adsorbed helium on krypton-plated
graphite, \cite{tfv}, (for a general reference see \cite{lawrye}).  

\smallskip
As is well known in the $d=2$ minimal 
conformal theories there exists a finite number
of primary operators \cite{bpz}.
In TIM they can be divided in two classes according to
their parity with respect to a natural $Z_2$ symmetry: the energy operators
are even and the spin operators are odd.
At the critical point there is a
spontaneous breaking of the $Z_2$ symmetry.
TIM  displays other symmetries apart from the conformal one:
superconformal symmetry \cite{fqs,qiu}
and those based on the coset formulation
($su(2)$ and $e_7$ algebras, see  \cite{go}).

\smallskip
In this article 
we  study  the two point
correlation functions of the energy operator and of  
the spin operator in proximity of the tricritical point.
 More specifically
we  consider the model obtained by adding
to the conformal action the most relevant energy operator
of the critical theory, $\e$ 
(of scale dimension $1/5$, see Section \ref{timtim}):
\beq
A= A_{CFT}+\lambda \int d^2 x \e (x).
\eeq
We also restrict ourselves to
 the high temperature phase of the model,
that corresponds to positive values of 
the coupling $\lambda$.

The considered  model is integrable 
and 
its spectrum and $S$ matrix are known, \cite{cm}
(see also \cite{integrable}).
 In the 
high temperature phase the massive excitations are given
by seven elementary particles. 
The mass of the fundamental particle is
given by:
\beq\label{cfateev}
m=\cf \lambda^{5/9},
\eeq
where $\cf$ is \cite{fateev}:
\bea
\cf&=&
\left({2{\Gamma({2\over9})}\over{\Gamma({2\over3})}{\Gamma({5\over9})}}\right)
\left({4\pi^2\Gamma({2\over5}){\Gamma({4\over5})}^3\over
{\Gamma({1\over 5})}^3{\Gamma({3\over 5})}
} \right)^{5/18}\non\\
&=&
3.745372836243954223      \cdots
\eea
The masses of the other particles can be found in \cite{amv} with
a more detailed analysis of the model. 
In the same paper the authors were able to find the form factors
of the model (see \cite{ff} for general references
and \cite{dm}
for recent developments)
 and to calculate the energy-energy correlation function
including the excited states given in table 3 of that work.
The representation of the correlation function in terms of form 
factors gives rise to
a long distance expansion 
(the small parameter being $e^{-mr }$)
that is
well appropriate in 
the region down to $mr \simeq 1$, but is reasonably 
less quickly convergent 
in the 
$mr << 1$ region.

\smallskip
In this paper we study the 
energy energy and spin spin correlators
by use of a perturbative expansion around the conformal theory:
in this case the small parameter is $mr$ and
consequently  our analysis 
is 
complementary to the
form factor approach.

The perturbative calculations are performed 
by using
the Operator Product Expansion
(OPE) 
approach
in order to overcome the
infrared divergences problem 
typical of the relevant perturbations of massless
theories.
The idea of this approach came out very early in the 
general context of
quantum field theories \cite{ideas}
(see \cite{qcd} for similar ideas in QCD),
was first introduced  in the context of 
perturbed conformal theories by Al. B.
Zamolodchikov in \cite{genio} 
(see also \cite{sonodate} for an analog independent
proposal 
and \cite{demu} for an early application of Ref.\cite{genio}), and was  
developed at all orders in a very general context
 in
\cite{guima1}.

In the OPE approach the terms non analytic in $\lambda$, 
whose very existence is related 
to  infrared divergences of naive perturbation 
theory, are  segregated
in the Vacuum Expectation Values  of the operators
present in the OPE expansion
of 
correlators
\beq\label{ope}
<\Phi_a (r) \Phi_b(0)>_\lambda \sim \sum_c 
C_{a b}^c(r, \lambda )<\Phi_c(0)>_\lambda
\eeq
($\Phi_a$ 
are deformations of the conformal theory operators,
 the suffix $\lambda$ refers to the complete theory correlators;
the dependence on $\lambda$ of the Wilson coefficients will be omitted
in the following).
The resulting perturbation theory for the
Wilson coefficient $C_{a b}^c(r, \lambda )$ 
turns out to be IR finite at all orders.
The  required operators VEV are a non perturbative
additional information that 
should be  considered as an external input in this context. 

In the specific
example that we consider, the  mean value 
of the energy operator $\varepsilon$ will be determined by 
the 
Thermodynamic
Bethe Ansatz, \cite{fateev},
and the mean value of the sub-leading
 energy operator, required for our purposes, 
will be estimated to a reasonable accuracy
 by using two exact sum rules
that express the scale dimension of the
energy 
and the 
central charge of the unperturbed
conformal theory in terms of moments of the
energy energy correlators
\cite{cardy,cadesi}
(in the long distance region
we use the results of \cite{amv} for the correlator).

Our final result is an estimate 
of the energy energy
and  spin spin 
correlators  up to $(mr)^2$. 
The prediction for the energy energy operator
 will then be compared with those of  \cite{amv}
in Figure 1. 

The plan of the paper is as follows: in Section \ref{timtim}
we collect the required knowledge on the TIM and introduce
the first order perturbative formulas for the
 Wilson coefficients; in Section \ref{results}
the final $O((mr)^2)$ expression for the correlators are 
reported; in Section \ref{sumrules} the missing  VEV
of sub-leading energy 
is estimated by use of exact sum rules
and long distance results.
Details of calculations are reported in \ref{integral}
while  \ref{theorem} reports a 
very general theorem that turned out to be useful
in calculations. 

\section{Tricritical Ising model}\label{timtim}
The Tricritical Ising Model (TIM) is 
 the minimal CFT with central
charge $c=\frac {7}{10}$ 
($\alpha _-^2 =\frac {4}{5}$, in notation of \cite{dot3}).
At the critical point we have the following operators:

$\bullet $the identity $\Phi _{11}=I , \Delta _I=0$,

$\bullet $the magnetization field $\Phi _{22}=\sigma  , \Delta _{\sigma } =\frac
{3}{80}$,

$\bullet $the sub-leading magnetization operator $\Phi _{12}=\alpha  , \Delta
_{\alpha } =\frac {7}{16} $,

$\bullet $the energy density $\Phi _{21}=\varepsilon  , \Delta _{\varepsilon }
=\frac {1}{10} $,

$\bullet $ the vacancy operator or sub-leading energy operator $\Phi _{31}=l , 
\Delta _l=\frac {3}{5}$,

$\bullet $ the irrelevant field $\Phi _{41}=\varepsilon '' , \Delta 
_{\varepsilon ''}=\frac {3}{2}$.

Using the fusion rules and knowing the Wilson coefficients we can construct the
algebra for TIM:

\begin{tabular}{|c|c|c|c|c|c|} 
\hline      $$& $\varepsilon ''$ &$\sigma $&$l$&$\alpha $&$\varepsilon $\\ 
\hline
$\varepsilon $ &$l$&$\alpha +\sigma $&$\varepsilon ''+\varepsilon $&$\sigma $&$I+l$
\\ 
\hline $\alpha$&$\alpha $&$\varepsilon +l$&$\sigma $&$I+\varepsilon ''$&$\sigma    $
 \\ 
\hline
$l$&$\varepsilon $&$\alpha +\sigma $&$I+l$&$\sigma $&$\varepsilon ''+\varepsilon $
 \\ 
\hline $\sigma $ &
$\sigma $&$I+\varepsilon +l+\varepsilon ''$&$\alpha +\sigma $&$\varepsilon +l$&
$\alpha +\sigma $
\\ \hline $\varepsilon '' $&
$I$&$\sigma $&$\varepsilon $&$\alpha $&$l$
 \\ \hline
\end{tabular}

\smallskip
The corresponding  Wilson coefficients can be found in 
Table 4 of \cite{lassig} (their normalization:
$<\Phi_i\Phi_j>=\delta_{ij}$ will be used everywhere here.)

Let us write here the 
adimensional Wilson coefficient   
$\widehat{C^l _{\e \e }}$ (${\widehat C}\equiv C 
|r|^{{\hbox {dim}} C}$):
\bea
\widehat{C^l _{\e \e }}
 &=& {2\over 3}\sqrt{{\Gamma (4/5) \Gamma ^3 (2/5) 
\over\Gamma(1/5)\Gamma^3 (3/5)}}\non\\
&=&0.6103020742648446931 \cdots
  \eea 
and
\beq
\widehat{C^\e _{\s \s }}={3\over 2} \widehat{C^l _{\e \e }}
\eeq

We consider in this paper the following perturbation:
\beq
A= A_{CFT}+\lambda \int d^2 x \e (x)
\eeq
where $\lambda\sim T-T_c >0$ and $\e$ is the most relevant 
energy operator of TIM, with total dimension $1/5$.

The idea of the OPE approach (see  \cite{ideas,genio,sonodate}) 
is to confine contributions non analytic
in $\lambda$ inside the VEV's
and to build up an IR finite 
perturbation theory (Taylor expansion around  $\lambda=0$)
for the Wilson coefficients.
Assuming the regularity of the Wilson coefficients in 
terms  of the 
renormalized coupling $\lambda$, 
the asymptotic convergence of OPE
and 
the validity of 
the Renormalized Action Principle, 
that in our case looks like 
\beq\label{actionprinciple}
-{\de \over \de\lambda}<[\cdots ]>_\lambda= \int <[:\e: \cdots] >_\lambda,
\;\; \;:\e: \equiv \e-<\e>_\lambda
\eeq
($[\cdots]$ means renormalization, when needed),
an IR finite representation
  for 
the  generic $n^{th}$ derivative of $C_{ab}^c$ with respect to
 the coupling,  evaluated at $\lambda=0$,
 has been given \cite{guima1},  involving  integrals of 
 conformal correlators.
See \cite{guima2} for an application to the 
critical Ising model perturbed by a magnetic field.

In our case we will 
 approximate the correlators by use of the
following first order expressions
(all derivatives below are intended to be 
evaluated  at $\lambda=0$):
\bea
<\e \e>_\lambda &=&C^1 _{\e \e}
+C^l _{\e \e}<l>_\lambda+\lambda \de _\lambda C^\e _{\e \e} <\e>_\lambda  +
O(\lambda ^2) \label{coree}
\\ 
<\sigma \sigma >_\lambda&=& C^1 _{\sigma \sigma }+\lambda \de _\lambda C^1_{\sigma \sigma}
+
C^{\e} _{\sigma \sigma} <\e >_\lambda+
\non\\
& &
\lambda \de_ \lambda C^{\e} _{\sigma \sigma}<\e >_\lambda+C^l _{\sigma \sigma}<l>_\lambda
+O(\lambda^{5/3}).\label{corss}\eea

At first order 
we derive the following nontrivial relations,
(see also \cite{genio, sonodate} for first order formulas):
\beq
-\de_\lambda C_{\s\s}^1=\int'd^2z<\e(z)(\s(r)\s(0)-C_{\s\s}^\e\e(0))>
\label{pippo}\eeq
\bea& &
-\de_\lambda C_{\s\s}^\e
<\e(\infty)\e(0)>= \non\\
& &\int'd^2z
<\e(\infty)\e(z)(\s(r)\s(0)-C_{\s\s}^1-C_{\s\s}^l l(0))>
\label{pluto}\eea
\bea& &
-\de_\lambda C_{\e\e}^\e
<\e(\infty)\e(0)>= \non\\
& &\int'd^2z
<\e(\infty)\e(z)(\e(r)\e(0)-C_{\e\e}^1-C_{\e\e}^l l(0))>
\label{paperino}\eea
where the prime means any (rotation invariant)
 IR regularization of the integrals.
We observe that the derivative 
of Wilson coefficients is obtained by adding up
a "naive perturbative term" (insertion of the interaction $\int \e$
in the correlator)
and some  "IR counterterms" generated by the 
theory itself in a natural way: both
contributions are IR divergent 
but their sum is finite and gives the wanted result.
This structure persists in the 
general all order formulas. 
The explicit form of the correlators and the 
details of the calculation of the IR finite part 
are reported in \ref{integral}.

We can parameterize the VEV of  $\e$ and $l$ as 
\bea
 <\e>_\lambda&=& A_{\e} \lambda^{1/9}\\
<l>_\lambda&=& A_{l} \lambda^{2/3} \label{veval}
.\eea
Notice that potential adimensional
logarithms are absent in the VEV's 
because no renormalization 
effects arises in this case (an operator of 
dimension $x$ can mix in this theory only 
with operators of dimension $x-{9\over 5} k$
for positive $k$).

The constant $A_\e$ can 
be fixed by use of results of \cite{fateev}
(Thermodynamic Bethe Ansatz)
\bea
A_{\e}
&=&-{5\over 36} {S({2\over9})\over S({1\over3})S({5\over9})}
\left({4\pi^2\Gamma({2\over5}){\Gamma({4\over5})}^3\over
{\Gamma({1\over 5})}^3{\Gamma({3\over 5})}
} \right)^{5/9}
\left({2{\Gamma({2\over9})}\over{\Gamma({2\over3})}{\Gamma({5\over9})}}\right)^2
\\
&=&
-1.468395424027621489 \cdots
\eea
where $S(x)=\sin\pi x$
(see e.g. \cite{guima2} for more details).
The quest for an estimate of $A_l$ is the goal of Section \ref{sumrules}

\section{Results}\label{results}
We report in this Section the results for the approximate
energy energy and spin spin correlators, Eqs.(\ref{coree}-\ref{corss}).
More details on the calculations can be found in \ref{integral}.
 
It is convenient 
to define adimensional quantities ${\widehat C}\equiv C 
|r|^{{\hbox {dim}} C}$, 
$F_{A,B}\equiv {{<AB>_h}}|r|^{{\hbox {dim}} AB}$
and introduce the scaling variable
$\tau =mr$, where $m$ is the fundamental mass of the model, defined in 
Eq.(\ref{cfateev}). 

We have:
\bea
F_{\e\e}&=& 1+A_l{\widehat C_{\e \e}^l} \tcf^{6/5}+
A_\e\widehat{\de _\lambda C^\e _{\e \e }} \tcf^{2}+O(\tau^{18/5})\label{ee}
 \\
F_{\s \s}&=& 
1+A_{\e} \widehat{C}^{\e}_{\s \s} \tcf ^{1/5}
+ {A}_l \widehat{C}^l _{\s \s}\tcf ^{6/5}\non\\
& &
+\widehat{\de _\lambda C^1 _{\s \s }}\tcf ^{9/5}
+{A}_{\e} \widehat{\de _{\lambda} C^{\e}_{\s \s}}
\tcf^2+O(\tau^3)
\label{ss}\eea

By use of results in \ref{integral} we obtain:
\bea
-\widehat{\de_\lambda C_{\e\e}^\e}
&=&\pi^3
     {{\Gamma({2\over 5})}^2{\Gamma({7\over 10})}^2 \over 
{S({1\over 5})}^2{S({2\over 5})}^2S({6\over5})
S({14 \over 5})}
{(S({4\over5}) - S({8\over 5}) - S({16\over 5}))\over 
2^{6/5}{\Gamma({1\over5})}^4{\Gamma({3\over 5})}^4}                                            \non\\
&=&-1.008826378183031     \cdots\\
-\widehat{\de_\lambda C_{\s\s}^1}&=&
-{{S({1\over 10})}^2\over S({1\over 5})}
{{\Gamma ({9\over10})}^4\over{\Gamma({9\over 5})}^2} 
\left({\Gamma({4\over5}) {\Gamma({2\over 5})}^3  \over
\Gamma({1\over5}){\Gamma({3\over5})}^3 }    \right)^{{1\over2}}
\non\\
&=& -0.223579589302507
\cdots\\
-\widehat{\de_\lambda C_{\s\s}^\e}
&=& 25\pi {{\Gamma({3\over 5})}^2{\Gamma({13\over10})}^2S({1\over5})
{S({7\over 10})}^2
     {S({7\over5})}^2 \over {S({2\over5})}^2{S({4\over5})}^2
S({8\over5})S({23\over10})
     S({29\over10})             }
{(-S({4\over5}) + S({11\over5}) + S({12\over5}) + 
       S({18\over5}) - S({27\over5})) \over
2^{12/5}{\Gamma({1\over5})}^4}\non
\\&=&-0.266530272188332
 \cdots\label{mainresult}
\eea
All other quantities have been defined  previously;
for a numerical value  of $A_l$ see Eq.(\ref{al}).

\section{Sum rules} \label{sumrules}
In this section we will consider two known 
sum rules satisfied by 
the exact complete theory.

The sum rule more sensitive to the short distance 
behavior of the correlators is
the one concerning the scale dimension of operators.
If we consider, as is the case,
 a perturbation $\lambda \int\e$ of a conformal field theory
(assumed to draw the system towards a massive theory) 
and 
an operator $X$ with scale dimension
$\Delta_X$, 
assumed to have a VEV $<X>_\lambda=A_X \lambda^{\Delta_X /(1-\Delta_\e)}$,
then, by taking the derivative with respect to $\lambda$ of the VEV
and by
 using  the Action Principle hypothesis, 
Eq.(\ref{actionprinciple}), it is easy to 
obtain 
\beq
{\Delta_X\over 1-\Delta_\e}
=-\lambda \int d^2x <[X(x) \e(0)]>_{\lambda \; c}/<X>_\lambda
.\eeq
This exact sum rule 
has been derived first,  in a more  generalized form,
 in \cite{cadesi}. 

By choosing $X=\e$ and using the known relation 
between the trace of energy momentum tensor 
$\Theta$
and the perturbing field $\e$
(holding at least when $\Delta_\e<1/2$),
\beq
\Theta=4\pi \lambda (1-\Delta_{\e})\e
\eeq
we have the 
equivalent relation:
\beq\label{apsr}
\int  d^2x <\Theta \Theta >_{\lambda \; c} 
= -{36\pi^2\over 25}{A_\e\over \cf^2} m^2. 
\eeq

The second sum rule we consider
is given by 
\beq\label{csr}
c\equiv {7\over 10} = {3\over 4\pi}\int d^2 x |x|^2 
<\Theta \Theta >_{\lambda \; c}
\eeq
and has been obtained in the context of the C-theorem
\cite{cardy}.

The  short distance behavior 
of the connected $\Theta \Theta$ correlator can
be expressed in terms of the scaling function 
$F_{\e\e}$ of Eq.(\ref{ee}) as
\beq\label{tete}
<\Theta \Theta >_{\lambda \; c}
/m^4 =\left({18\pi\over 5\cf^2}\right)^2 \tcf^{-2/5}
\left(F_{\e\e}-A_\e^2 \tcf^{2/5}\right) 
\eeq
(Notice that we subtracted $<\e>_\lambda^2$ to get the connected correlator). 

The long distance behavior is given by:
\beq\label{ldtete}
<\Theta \Theta >_{\lambda \; c}
 /m^4 \sim 
{1\over\pi} (f_2 ^2 K_0 (c_2 \tau  ) + f_4 ^2 K_0 (c_4 \tau )) 
\eeq
where $c_2=2\cos (5\pi/18)$, $c_4=2\cos (\pi/18)$ 
and 
$ f_2 = 0.9604936853$, $f_4 = -0.4500141924$, \cite{amv}.

We now proceed in the same way as done in \cite{guima2}, where
good results were obtained: 
we split the integrals in the sum rules 
at some point $mr=\Lambda$, evaluating
$mr<\Lambda$ region with the short distance expression 
Eqs.(\ref{ee}),(\ref{tete}) and $mr>\Lambda$  
with Eq.(\ref{ldtete}).
In particular the
 integral of the long distance correlator is given by
\beq
\int_{|mr|>\Lambda} <\Theta \Theta >_{\lambda \; c} d^2 \tau /m^4
 =2\Lambda(f_2 ^2 /c_2 K_1 (c_2 \Lambda)
+ f_4 ^2 /c_4 K_1 (c_4 \Lambda))
\eeq
while for the c sum rule the integral of the long distance 
contribution is given by
\beq
\int_{|mr|>\Lambda} <\Theta \Theta >_{\lambda \; c}  r^2 d^2 x = 2 (f_2 ^2 /c_2 ^4 \widehat{K}(c_2 \Lambda)+ 
f_4 ^2 /c_4 ^4
\widehat{K}(c_4 \Lambda))
\eeq
where
\beq
\widehat{K}(x) = (4 x+x^3)K_1 (x) +2 x^2 K_0 (x).
\eeq
By imposing the sum rule Eq.(\ref{apsr})
we can recover the unknown constant
$A_l$ as a function of 
$\Lambda$. 
Following \cite{guima2} a
 good approximation for $A_l$ can be obtained 
by looking
for a minimal sensitivity point $\Lambda^*$ of $A_l$.
We obtain in this way
\beq\label{al}
A_l=3.78 
\eeq
where an  error of $5-10\%$ is extimated by using  
different variants of the method
(imposing validity of Eq.(\ref{csr}) or 
approximate validity of both).

\section{Conclusions}
By use of results of Sections \ref{results}-\ref{sumrules},
we can give a prediction on the short distance behavior
of the energy energy (or $\Theta \Theta $) and spin spin 
correlators 
of the  TIM model near criticality
(most relevant energy perturbation).
As a by product of our work we get an estimate
for the subleading energy VEV, Eqs.(\ref{veval})-(\ref{al})
                         
The most striking feature of OPE approach used in this paper,
 is 
the fact that non analyticities in $\lambda$,
present in the short distance behavior of correlators,
are naturally recovered from the VEV and manifest
in non negligible fractionary power corrections (in $\lambda$)
in  
Eqs.(\ref{ee}) - (\ref{ss}).

In Figure 1 we compare our predictions for 
the $\Theta \Theta$ correlator as a function 
of $mr$
with the corresponding long distance 
prediction of \cite{amv}.
The agreement between the two 
approaches
is  good in the intermediate 
$0.5<mr<1.5$ region,
where corresponding "small" parameters are actually
of order $1$!
See \cite{guima2,zamo,dm} for similar situations. 
Moreover the lower $mr$ region shows a 
 reasonable evidence of convergence
of the long distance expansions towards our results.
We remark in particular,
 referring also 
 to an similar comparison 
made by the authors for the critical Ising Model in magnetic
field \cite[fig.2]{guima2},
that the weaker the singularity in $r=0$ 
of a correlator is,
the fastest is the convergence of the 
long distance approximants.
We think that, in any case, 
the two approaches
are always complementary.

Finally we want to stress that the OPE approach
gives an all order IR finite perturbation theory that 
is very general 
(mild and reasonable hypotheses are behind it) 
and do not uses integrability essentially:
the computation of Wilson coefficients 
can thus reasonably
applied to many others physically interesting
models in statistical mechanics.
The main problem of reaching higher orders is 
related to the complexity of the required integrals.
We think nevertheless that second order expressions
when necessary, could be reached, at least numerically.
 The ignorance of the expressions of the 
VEV will result
at each finite order
in a finite
set of {\it universal} constant parameterising 
all the correlators. 
How to reach a non-perturbative  information
on those constants, apart from what
already known  
from Thermodynamic Bethe Ansatz, is 
still an open problem.

{\bf Acknowledgements:}
The authors want to thanks Calin Buia, who 
participated at the early stages of the work,
J.B. Zuber  for reading the manuscript,
and G. Mussardo and C. Acerbi for giving them a table of
numerical  data.
R.G. thanks H. Navelet, R. Peschanski, M. Bauer for stimulating discussions
on the theorem in \ref{theorem}.
Part of this work has been done during a visit of the authors to CERN,
whose warm hospitality is acknowledged.
The work of R.G. is supported by a TMR EC grant, contract N$^o$ ERB-FMBI-CT-95.0130.
R.G. also thanks  the INFN group of Genova for the kind hospitality.

\appendix
\section{A useful theorem}\label{theorem}
We give here a (slight generalisation of) a useful theorem
proved in \cite{mathur}, (see also \cite{dot3,const} for related work).

Let us consider a  double integral on the plane of the 
form
\beq\label{forma}
I=\int d^2w \sum_{\alpha ,\beta=1}^{N} f_\alpha(w) 
Q_{\alpha \beta} \bar{f}_\beta(w^*)
\eeq
where $\{f_\alpha(w)\}_{\alpha=1}^N$ 
$\{\bar{f}_\beta(w^*)\}_{\beta=1}^N$ are two sets of 
linear independent functions,
 $Q_{\alpha \beta}$ is a complex, constant, matrix
and ${}^*$ denotes the complex conjugate.


We will assume that $\bar{f}_\beta(w^*)$ and 
$({f}_\beta(w ))^*$ have the same monodromy structure.
In particular 
we assume firstly that 
the two sets of functions
 $f_\alpha(w)$ and $g_\beta(w)\equiv(\bar{f}_\beta(w^*))^*$
have the
same (algebraic) branch points, $\{w_k\}_{k=0}^{m+1}$, such
\beq
0=|w_0|<|w_1|<\cdots<|w_m|<|w_{m+1}|=\infty
\eeq 
and that are analytic elsewhere.
(The relative cuts in the $w$ plane are 
taken to stay along the
same line $\gamma$ 
connecting the $w_k$'s and chosen such that $|w|$ is increasing along it.
Cuts in $w^*$ plane are obtained by ${}^*$ mapping the cut $w$
plane.) 
In the following we will require also  that
for each choice of $k=0,\cdots, m+1$,
in the limit $w\rightarrow w_k$, 
 \bea
|(w-w_k) f_\alpha(w)|&\rightarrow&0 \non\\
|(w-w_k) \bar{f}_\beta(w^*)|&\rightarrow&0 ,
\eea
uniformly on the cut plane. This condition guarantees the
convergence of the integral  Eq.(\ref{forma}) and the validity 
of some steps of the proof.

Secondly we suppose  that $f$ and $g$ have the
same
related monodromy matrices $M_k$,
 defined by 
the requirement
\beq
f(w_-)= M_k f(w_+).  
\eeq
$M_k$ above is  the same constant matrix for {\it all}
 points $w$  staying in the region of $\gamma$ between 
$w_{k-1}$
and  $w_{k}$ and ${}_+$ (${}_-$) means limit from above (below)
the cut
(similar definition for $g(w)$). 
In particular $w_\pm$ are connected by a
path in the cut plane enclosing all $w_j$ with $j<k$.

We assume finally  that  the matrix $Q$ 
is invariant by the monodromy group action:
\beq
\forall k \;\;\; Q=M_k^{t}QM_k^{*}
.\eeq 

All requirements together enforce the 
single valuedness of
the integrand and the stronger relation
\beq
f(w'_-) Q {\bar f}( w_-^* )= 
f(w'_+ ) Q {\bar f}( {w_+}^* )
\label{strong}
\eeq
for any  $|w|,|w'|$ on 
the region of $\gamma$ between 
$|w_{k}|$ and $|w_{k+1}|$ (the same for both).

It follows then, after a simple generalisation of the 
proof of \cite{mathur} based on Stokes' theorem
(and using strongly Eq.(\ref{strong})),
that we can express $I$ in terms of
one dimensional integrals as
\bea
I&=&{i\over 2}\sum_{k=1}^{m}
{\cal I}^{(k)}_\alpha
\left[ \left((1-M_{k+1})^{-1}-(1-M_k)^{-1}\right)^{t}  Q
\right]_{\alpha \beta} 
{\bar{\cal I}}^{(k)}_\beta
\label{niceformula}
\eea
where, ${}^t$ means transposition,
\bea
{\cal I}^{(k)}&\equiv& \int_{C_k} dw f(w) \\
{\bar{\cal I}}^{(k)}&\equiv&\int_{{\bar C}_k} dw^* {\bar f}(w^*)
\eea
and $C_k$ (${\bar C}_k$) are counter-clockwise (clockwise)
circumferences
starting at   
${w_k}_+$ (${w_k}_+^*$) 
and ending at 
${w_k}_-$ (${w_k}_-^*$),
 enclosing all the $w_j$ ($w_j^*$) of lower modulus.
We refer to \cite{mathur} 
for more details (see also  \cite{navelet} for 
related  integrals).

\section{More details on computations}\label{integral}
The expression for  the correlators
in Eqs.(\ref{pippo}-\ref{paperino})
can be obtained from the results of \cite{dot3}
for the second order correlators
of minimal conformal theories.
In particular the general structure is 
\bea &&
<\phi_{{ \mu }_1}(z_1)\phi_{{ \mu}_2}(z_2)\phi_{{ \mu }_3}(z_3)
\phi_{{\mu }_4}(z_4) > \propto\non\\
&&\prod_{i<j} |z_{ij}|^{2\gamma_{ij}}
\left(
S(a+b+c)S(b)|I_1(a,b,c;\eta)|^2
+S(a)S(c)|I_2(a,b,c;\eta)|^2
\right)\non
\eea
where the ordered multi-index
${\mu}=(n,m)$
labels operators and fixes 
their dimensionality 
(in our case $\e$ corresponds to 
$\mu=(2,1)$ while $\s$ to $(2,2)$),
and
\bea
S(x)&\equiv& \sin(\pi x)\non\\
\eta&\equiv&{z_{12}z_{34}\over z_{13}z_{24}}\non\\
I_1(a,b,c;\eta)&\equiv&
{\Gamma(-a-b-c-1)\Gamma(b+1)\over\Gamma(-a-c)}
{}_2F_1(-c,-a-b-c-1;-a-c;\eta)\non\\
I_2(a,b,c;\eta)&\equiv&
\eta^{1+a+c}
{\Gamma(a+1)\Gamma(c+1)\over\Gamma(a+c+2)}
{}_2F_1(-b,a+1;a+c+2;\eta)
.\eea
The relation between the multi-indices ${\bf\mu}$
and exponents $\gamma_{ij}$ and parameters
$a,b,c$ can be found in \cite{dot3} and 
will not be reported here. We denote with 
$
{}_pF_q(a_1,\cdots,a_p;b_1,,\cdots,b_q;z)
\equiv\sum_{k=0}^{\infty} {(a_1)_k\cdots (a_p)_k\over k!
 (b_1)_k \cdots (b_q)_k} z^k
$ the (generalised) hypergeometric function
(where $(a)_k\equiv\Gamma(a+k)/\Gamma(a)$).

The integrals we are interested in are
 therefore of the general form:
\bea
Z(a,b,c,d,e)&\equiv&
\int d^2\eta d^2z |z|^{2a} |1-z|^{2b} |\eta-z|^{2c}
|\eta |^{2d} |1-\eta |^{2e}  
\non\\
&= &
S(a+c)^{-1}\int d^2\eta |\eta|^{2d} |1-\eta|^{2e}\times  \non\\
& &
(S(a+b+c)S(b)|I_1(a,b,c,\eta)|^2+S(a)S(c)|I_2(a,b,c,\eta)|^2).
\non
\eea
The second line can be obtained by use of results of \cite{dot3}.
We evaluated the second part of the integral by use of the 
theorem reported in \ref{theorem}.
After the completion of the computation we realized that
an equivalent integral 
could be found  in Appendix
D of \cite{dot5} and we checked that the results are the same 
(modulo 
some non trivial algebraic relations satisfied by the 
generalised hypergeometric functions, \cite{pbm}). 
We report our expression 
that, by construction, involves a fewer number of boundary integrals 
(generalised hypergeometrics) to be computed.

Applying the theorem of \ref{theorem} to our needs,
 it is easy to realize that 
in
Eq.(\ref{niceformula}) there is only a contribution coming from 
${\cal I}^{(1)}$. 
If we pose 
\beq
{\cal I}^{(1)}=2 i(e^{i \pi d}S(d)J_1,e^{i \pi (a+c+d)}S(a+c+d) J_2)
\eeq 
 and compute the inverse of the  monodromy matrix terms, it follows after some algebra that:
\beq Z=z_{11} J_1^2+z_{22} J_2^2+z_{12}J_1J_2 \eeq
$$J_1\equiv \int_0^1z^d (1-z)^e I_1(a,b,c,z) $$
$$J_2\equiv \int_0^1z^d (1-z)^e I_2(a,b,c,z) $$

where  $S^{-1}(x)=1/S(x)$
and 
$$
J_1=B(b+1,-a-b-c-1)B(d+1,e+1) 
{}_3F_{2}(-c,-a-b-c-1,d+1;-a-c,2+d+e;1)
$$
$$
J_2=B(a+1,c+1)B(2+a+c+d,e+1)
{}_3F_{2}(-b,a+1,2+a+c+d;a+c+2,3+a+c+d+e;1)
$$

\bea
z_{11}& =&
-{1\over 4} S(a+c)^{-2} S^{-1}(c + d + e)S^{-1}(a + b + c + d + e)
\times\non\\
& &
      S(b)S(a + b + c)S(d) \times\non\\                   
& &
      ( S(b - c - d) - S(b + c - d) + S(b + c + d) - 
        S(2 a + b + c + d)   \non\\
& &
 - S(b + c + d + 2 e) + 
        S(2 a + b + 3 c + d + 2 e))
\eea
\bea 
z_{22} &=& {1\over 4}  S(a+c)^{-2}
S^{-1}(c + d + e)S^{-1}(a + b + c + d + e)\times\non\\
& &
     S(a)S(c)S(a + c + d)\times\non\\
& & 
    (S(a + b - d) + S(a - b + d) - S(a + b + d) + 
       S(a + b + 2 c + d) \non\\
& &
 - S(a - b - d - 2 e) - 
       S(a + b + 2 c +  d + 2 e))
\eea
\bea 
z_{12}&  =&  2 S(a+c)^{-2}
S^{-1}(c + d + e)S^{-1}(a + b + c + d + e)
     \times  \non\\ 
 & &
  S(a)S(b)S(c)S(a + b + c)S(d) S(a + c + d)
\eea
(We assumed the parameters to be real as it is the case of interest).

With the knowledge of $Z(a,b,c,d,e)$ 
the IR finite
part  of the regularized "naive" perturbative 
integrals can be extracted by use of Mellin transform techniques
without computing all IR counterterm integrals,
as explained in detail in \cite{guima2}, Section 2-3.
Using some known relations of ${}_3F_2$ hypergeometric functions
of argument $z=1$, \cite{pbm},
and fixing the normalisation of correlators by imposing
clustering properties (see \cite{dot6}),
 Eq. (\ref{mainresult})
can be recovered.

\bigskip

{\bf Figure Captions:}

{Fig 1:} Different estimates of $<\Theta \Theta>_{\lambda\; c}$ correlator
plotted vs $mr$: continuous curve is the short distance
approach (this paper);
 dashed curve refers
 to first two states  
 while dots correspond to the 
first $11$ states in form factor approach, \cite{amv}.

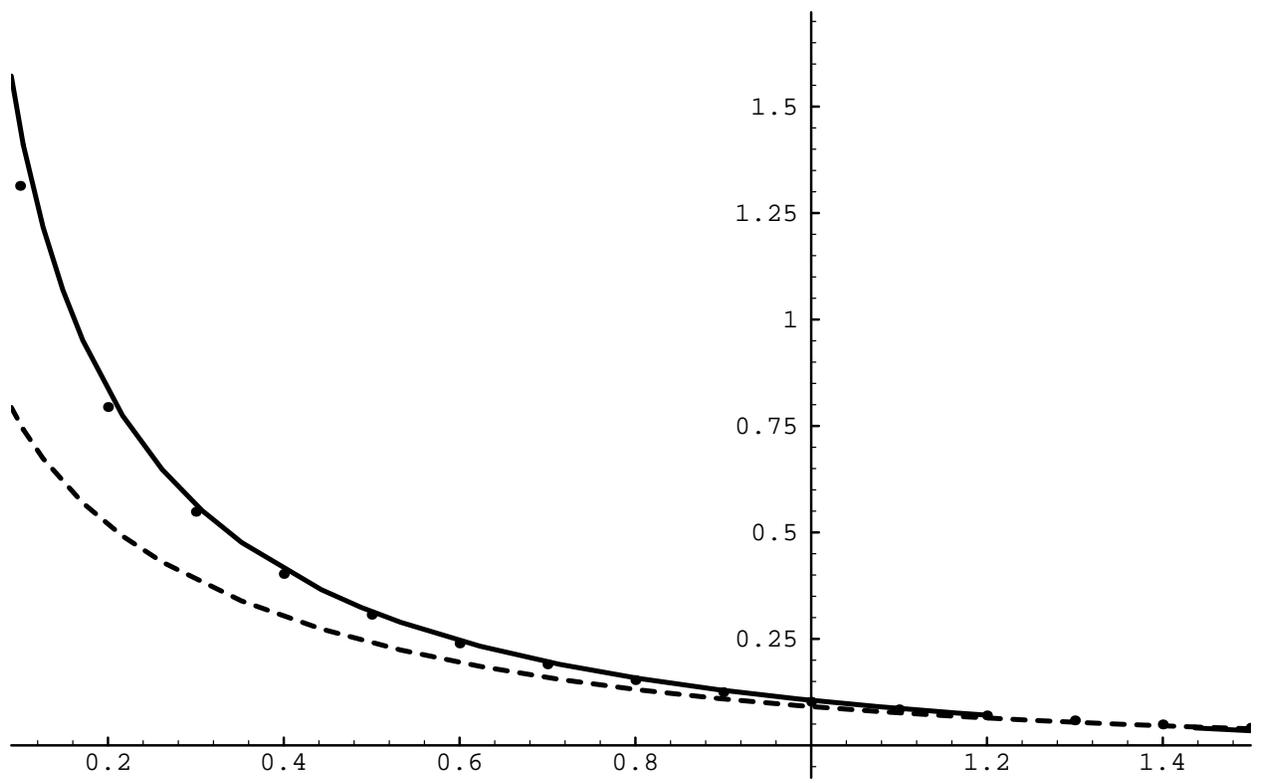
\end{document}